\documentclass[%
 aip,
 amsmath,amssymb,
 reprint,%
]{revtex4-1}

\usepackage{graphicx}
\usepackage{dcolumn}
\usepackage{bm}

\usepackage[utf8]{inputenc}
\usepackage[T1]{fontenc}
\usepackage{mathptmx}
\usepackage{etoolbox}

\makeatletter
\def\@email#1#2{%
 \endgroup
 \patchcmd{\titleblock@produce}
  {\frontmatter@RRAPformat}
  {\frontmatter@RRAPformat{\produce@RRAP{*#1\href{mailto:#2}{#2}}}\frontmatter@RRAPformat}
  {}{}
}%
\makeatother
\begin{document}

\preprint{AIP/123-QED}

\title[Articles]{Cascaded periodically poled electro-optical crystal optical phased array}
\author{Jingwei.Li}

\author{Yuchen.He}

\author{Huaibin.Zheng*}

\author{Sheng.Luo}

\author{Xin.Liu}

\author{Qingyuan.Hu}
 \email{huaibinzheng@xjtu.edu.cn}
\affiliation{Electronic Materials Research Laboratory, Key Laboratory of the Ministry of Education and International Center for Dielectric Research, School of Electronic and Information Engineering, Xi'an Jiaotong University, Xi'an, 710049, Shaanxi, China.}%

\author{Huaixi.Chen}

\author{Wanguo.Liang}

\affiliation{%
Fujian Institute of Research on the Structure of Matter, Chinese Academy of Sciences, Fuzhou, 350002, Fujian, China.
}%

\affiliation{%
Fujian Science \& Technology Innovation Laboratory for Optoelectronic Information of China, Fuzhou, 350108, Fujian, China.
}%

\author{Jianbin.Liu}

\author{Hui.Chen}

\affiliation{Electronic Materials Research Laboratory, Key Laboratory of the Ministry of Education and International Center for Dielectric Research, School of Electronic and Information Engineering, Xi'an Jiaotong University, Xi'an, 710049, Shaanxi, China.
}%

\author{Yu.Zhou}

\affiliation{MOE Key Laboratory for Nonequilibrium Synthesis and Modulation of Condensed Matter, Department of Applied Physics, Xi'an Jiaotong University, Xi'an, 710049, Shaanxi, China.
}%

\author{Xiaoyong.Wei}

\author{Zhuo.Xu}

\affiliation{Electronic Materials Research Laboratory, Key Laboratory of the Ministry of Education and International Center for Dielectric Research, School of Electronic and Information Engineering, Xi'an Jiaotong University, Xi'an, 710049, Shaanxi, China.
}%

\date{\today}

\begin{abstract}
Optical phased arrays (OPA) with high integration, fast speed, low power consumption, and high steering resolution are critical components in the emerging photonic integrated circuit (PIC), LiDAR, free space optical communication, 3D printing, and so on. According to the OPA working principle, its function is generally achieved by independently controlling the phase of the array elements. In practice, this presents a major challenge to overcome critical trade-offs of the element numbers, the control electronics, and the power consumption. Here, we give an alternative OPA solution to overcome this technical limitation, in the form of a cascaded periodically poled electro-optical crystal structure. Compared with the existing OPA scheme, only one control electronics is used to control the entire array elements in the current proposal, regardless of the number of array elements, implying higher integration and lower power consumption. With the help of the fast response properties of electro-optic crystal materials, a high-speed and high-resolution beam steering device is demonstrated. Simulating results show that the angular resolution can be improved by several orders of magnitude when the number of the cascaded-layer increases. An OPA prototype of a 6-layer cascaded periodically poled LiNbO$_3$ (cascaded-PPLN) was designed, fabricated, and characterized. The experiment that observed beam deflection in cascaded-PPLN OPA agrees well with the simulation results. Meantime, by demonstrating dynamic beam steering continually, its capability of continuous scanning and continually active phase tunability has been verified. Therefore, this cascaded periodically poled electro-optical crystal OPA offers a feasible direction of miniaturization and low power consumption for the optical system, such as the PIC system.
\end{abstract}
\maketitle

\section{\label{sec:level1}Introduction}

By controlling each emitter’s phase, integrated optical phased arrays (integrated-OPA) can tilt the wavefront of a light beam and enable nonmechanical beam steering\cite{bib1,bib2,bib3}, which are critical components in the emerging photonic integrated circuit (PIC)\cite{bib4,bib5,bib29}. They are also expected to be building blocks for emerging applications such as modern telecommunication networks\cite{bib6,bib7}, biomedical imaging\cite{bib8,bib9}, light detection and ranging (LiDAR)\cite{bib10,bib11,bib12}, and wireless optical communications\cite{bib13}. All of these applications require chip-scale OPAs that operate at low energy consumption, have ultrahigh electro-optic bandwidths and a precision aiming system, and feature no movable parts. Originally, an integrated-OPA platform based on electro-optical crystal, e.g. lithium niobate (LN), has been challenging to integrate on-chip because of difficulties in microstructuring crystal. Thereafter, many other photonic platforms compatible with microfabrication processes have emerged as attractive candidates, including those based on materials such as silicon\cite{bib1,bib14,bib15}, indium phosphide\cite{bib16,bib17}, polymers\cite{bib18,bib19}, and plasmonics\cite{bib20,bib21,bib22}. Although the integration problem has been greatly alleviated in these platforms, an integrated-OPA that meets these requirements simultaneously remains elusive because of the intrinsic limitations of the materials used. Recently, high-quality thin-film LN on insulator (LNOI) makes a breakthrough, which has paved the way for integrated LN photonics\cite{bib23,bib24,bib25}. However, the current research results show that the integrated-OPA scheme is still subjected to critical trade-offs of the emitters numbers, the control electronics and the power consumption\cite{bib1,bib26,bib27,bib28}, and thus is unable to reach the full potential of the material. The most OPAs exhibit the trade off between wide steering range and high resolution, making it difficult to achieve both a large steering angle and a narrow beam divergence. The resolution of OPAs is fundamentally tied to the number of array elements. This relationship can be easily understood as follows: the full-width at half-maximum (FWMH) divergence is determined by the equation ${{\psi }_{FWMH}}=0.886\cdot \lambda /Nd\cdot \cos \theta $, where $N$,$d$ and $\theta$ denote array number, array period, and steering angle, respectively, the steering angle $\theta =\Delta \varphi \cdot \lambda /2\pi \cdot d$, the resolution of OPAs is defined by $\theta /{{\psi }_{FWMH}}$. This implies the tradeoff between the steering range and angular resolution that can be improved only by increasing the number of control elements, which complicates the control system.  Whether it is possible to simultaneously achieve a precision high-speed beam steering system, an ultra-high bandwidth, a low energy consumption, and a high resolution in LNOI-OPA have remained a major challenge.

Here we demonstrate a feasible scheme, in the form of cascaded periodically poled electro-optical crystal for OPA, that overcomes the above trade-offs, featuring all the phase tuning units being manipulated by only one control electronics. The number of phase tuning units is related to the cascading layer. This structure is fabricated in LN crystal by designing the ferroelectric domain inside. The equal phase difference distribution between adjacent array elements (the slope of the near filed phase profile) is achieved after modulating by the electro-optical effect (Pockels effect). This cascaded periodically-poled OPA structure is robust and can achieve large angle, continuous and high-speed beam steering, which could be an outstanding candidate to achieve highly integrated and low power consumption PIC component. 

\section{Results and discussion}

\subsection{Principle of cascaded-PPLN OPA and simulation}

\begin{figure*}
\includegraphics[width=0.7\textwidth]{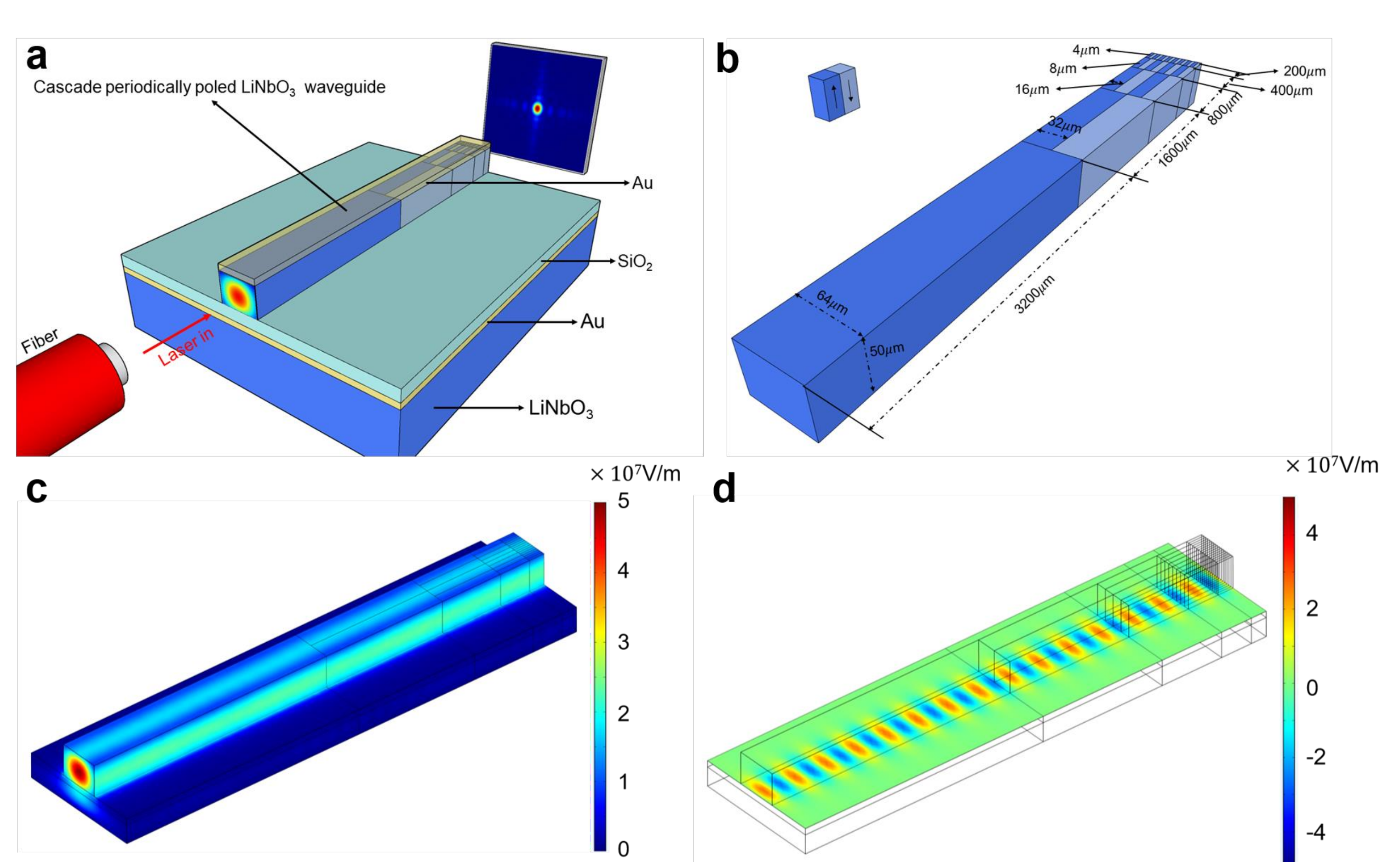}
\caption{\label{1}\textbf{a}, The overall layout of the on-chip cascaded-PPLN OPA. \textbf{b}, The schematic view of the waveguide core dimensions, dark blue for positive ferroelectric domains, light blue for negative ferroelectric domains. \textbf{c}, The optical mode profile at the wavelength of 1064 nm. \textbf{d}, The numerically simulated electric field distribution in the x-direction at the bottom of the waveguide.}
\end{figure*}

\begin{figure*}
\includegraphics[width=0.7\textwidth]{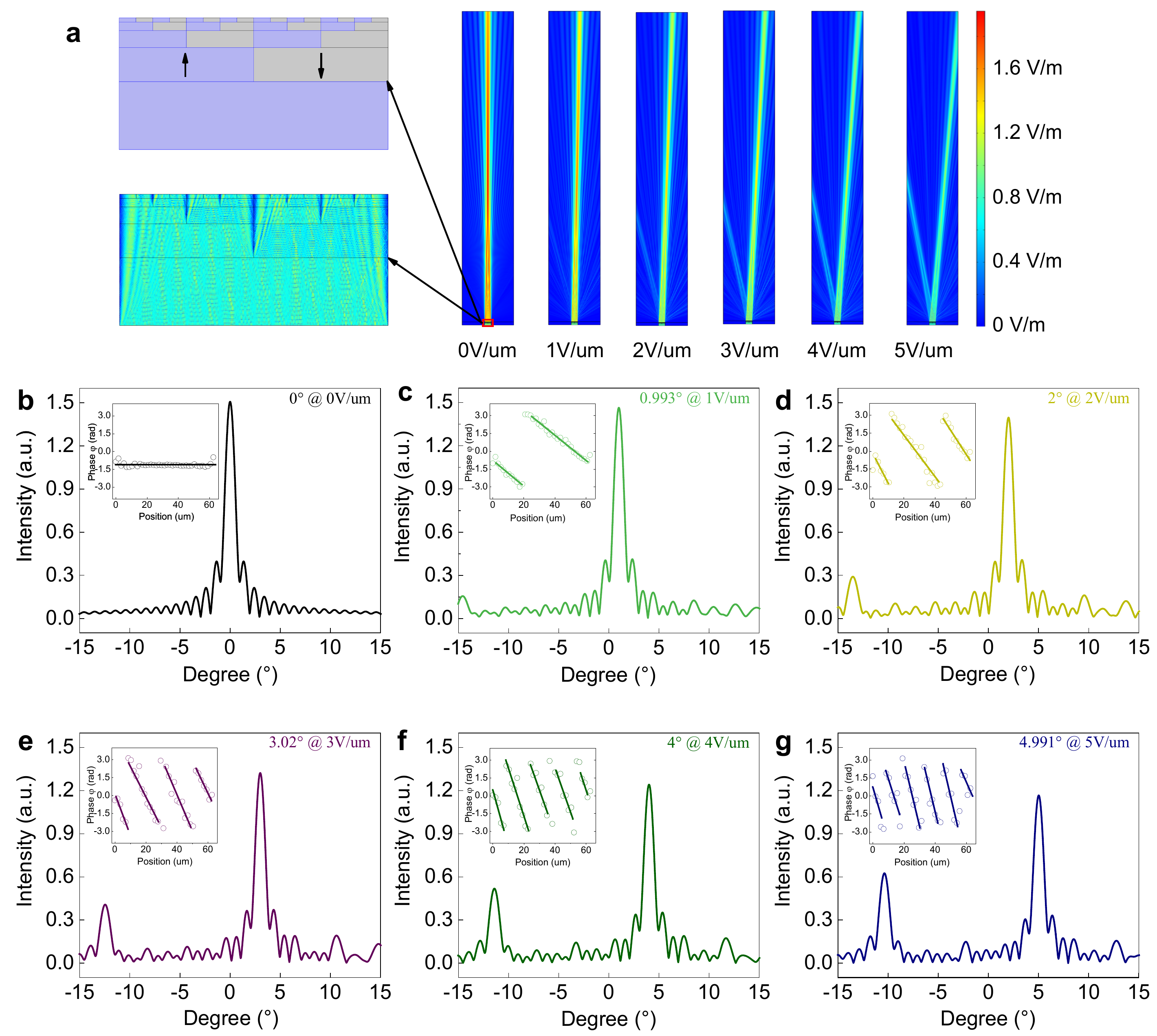}
\caption{\label{2}\textbf{a}, The electric field distribution of the cascaded-PPLN OPA waveguide at different voltage. The enlarge parts are the geometry of the cascaded-PPLN OPA waveguide and the electric field distribution inside the waveguide, respectively. \textbf{b-g}, The far field patterns of the cascaded-PPLN OPA waveguide at 0, 1V/$\mu$m, 2V/$\mu$m, 3V/$\mu$m, 4V/$\mu$m and 5V/$\mu$m , respectively, the inset shows the phase distribution curve of the electric field when the beam exits the waveguide.}
\end{figure*}

\begin{figure*}
\includegraphics[width=0.7\textwidth]{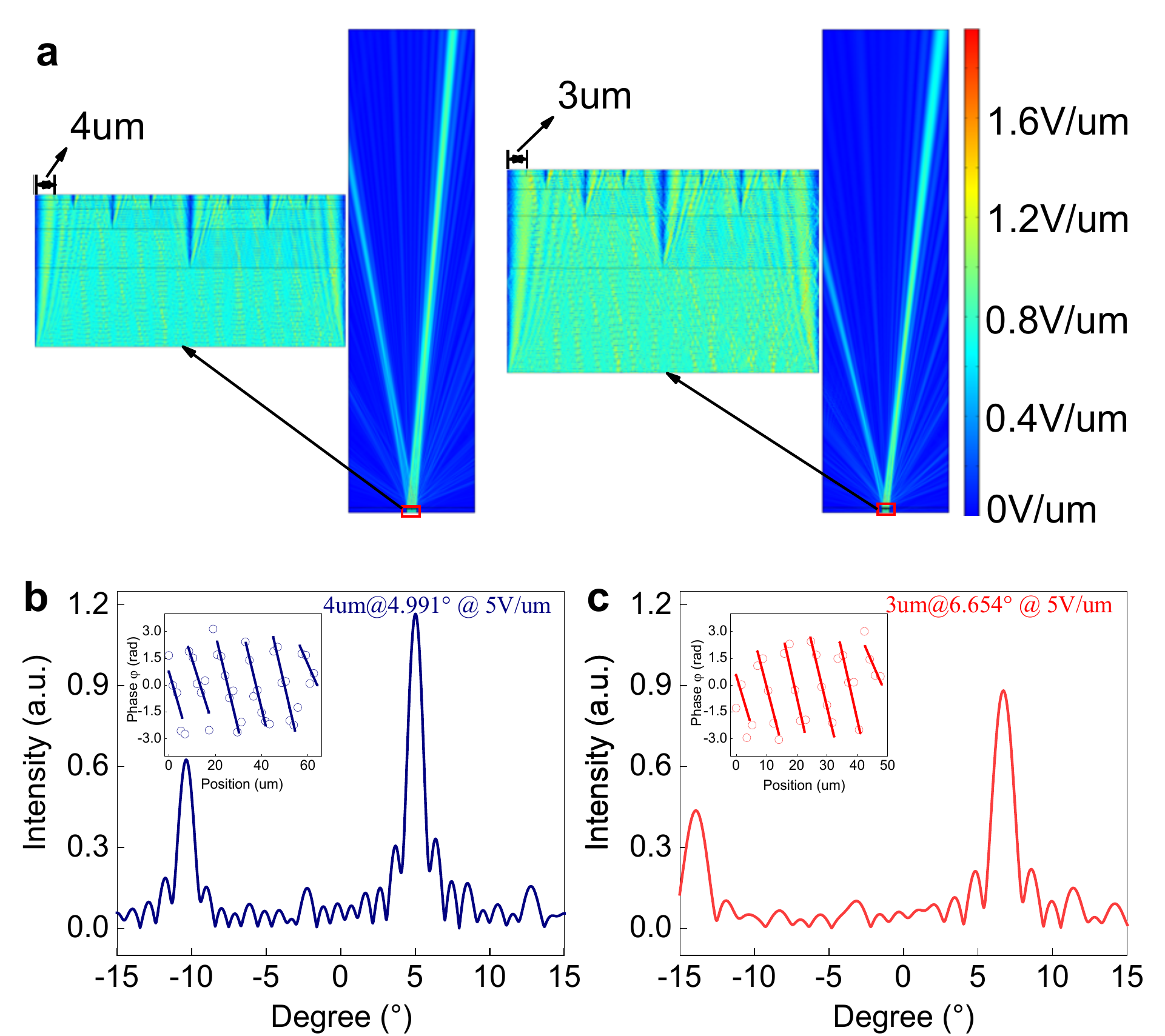}
\caption{\label{3}\textbf{a}, The electric field distribution of the cascaded-PPLN OPA waveguide at a voltage of 5V/$\mu$m when the widths of the individual electric domains in the last layer are 3 $\mu$m  and 4 $\mu$m, respectively, the enlarge parts are the electric field distribution inside the waveguide. \textbf{b and c}, The far field patterns of the cascaded-PPLN OPA waveguide at a voltage of 5V/$\mu$m with the widths of the individual electric domains in the last layer are 3 $\mu$m  and 4 $\mu$m, respectively, the inset shows the phase distribution curve of the electric field when the beam exits the waveguide.}
\end{figure*}

\begin{figure}
\includegraphics[width=0.4\textwidth]{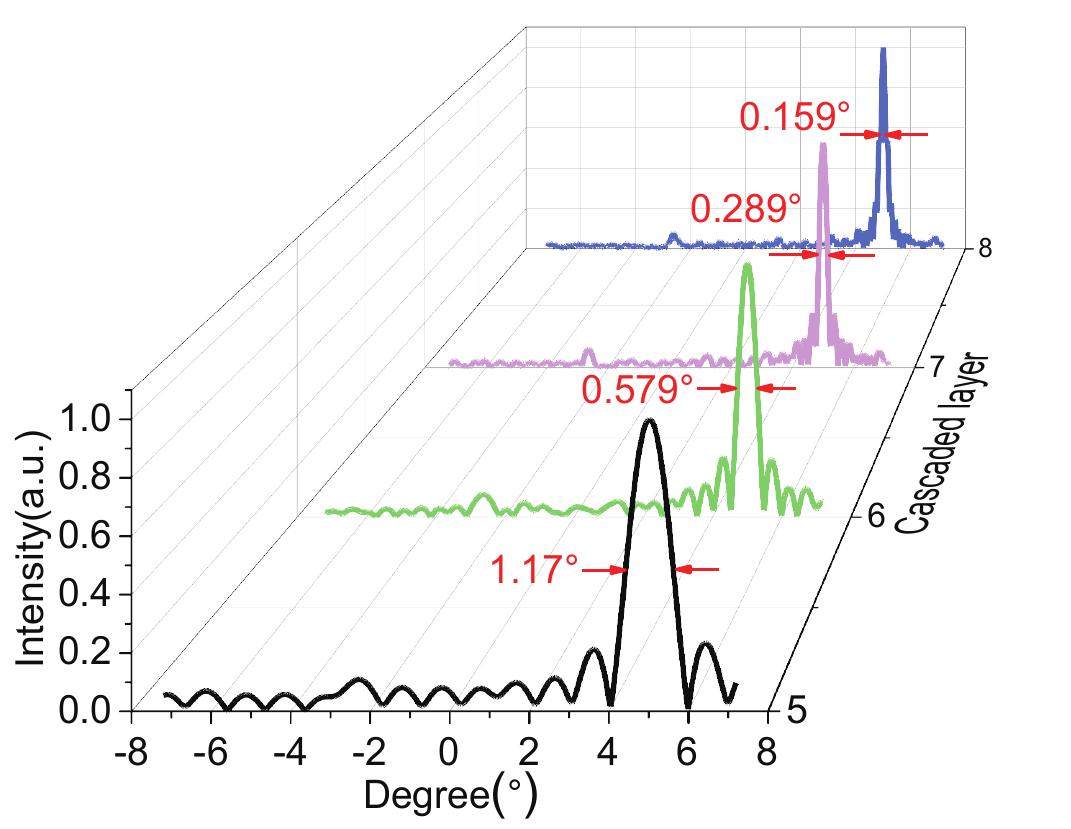}
\caption{\label{4} Far-field profiles for different cascaded layers PPLN OPA waveguide (5,6,7 and 8 cascaded-layer respectively) }
\end{figure}

\begin{figure*}
\includegraphics[width=0.7\textwidth]{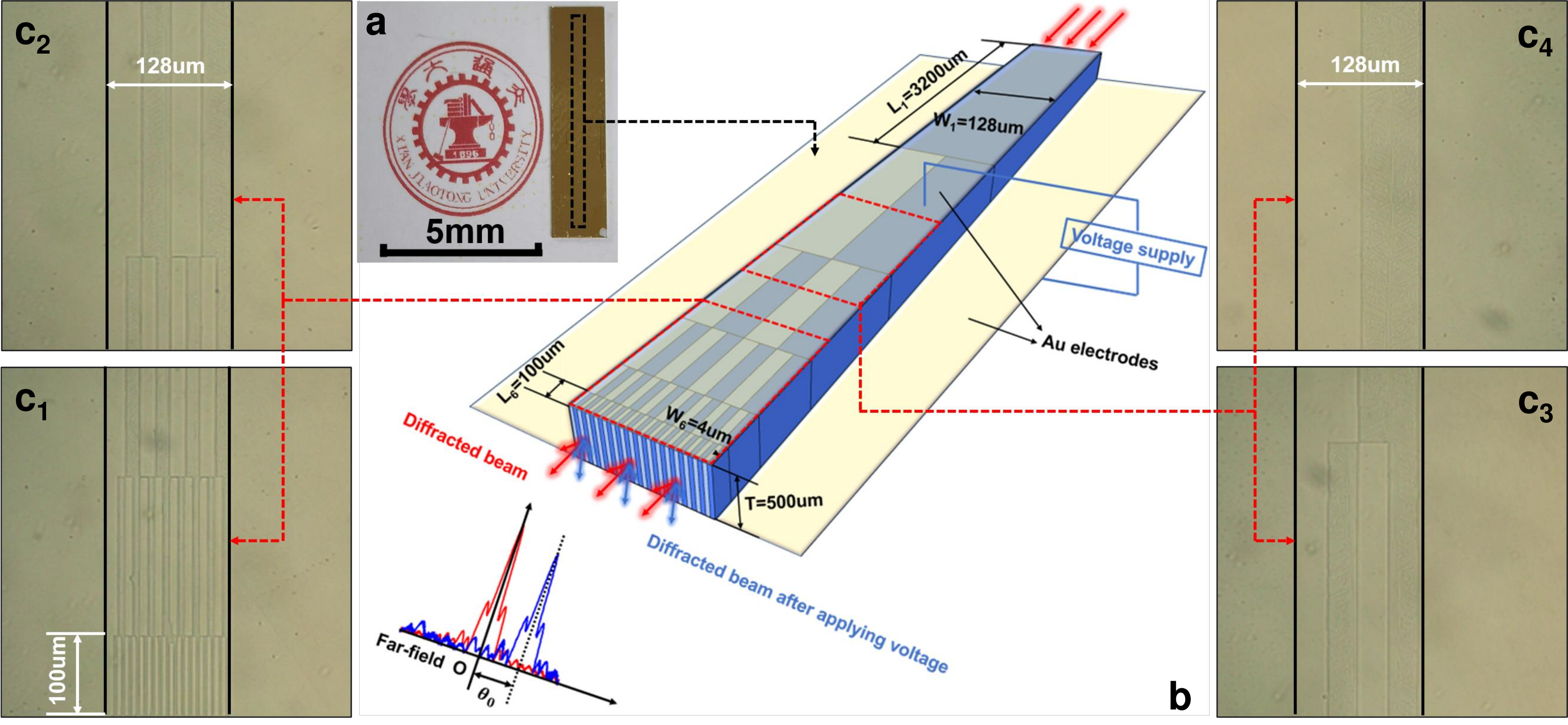}
\caption{\label{5}\textbf{a}, The 6-layer cascaded-PPLN OPA crystal with gold-plated electrodes. \textbf{b}, The geometry schematic of 6-layer  cascaded-PPLN OPA crystal crystal. The voltage is applied to the 6-layer  cascaded-PPLN OPA crystal along the Z-direction of the crystal. The beam is incident from the first cascaded layer and exits from the last cascaded layer. \textbf{c$_1$, c$_2$, c$_3$, c$_4$}, The SEM images of ferroelectric domains in different cascaded layers in 6-layer  cascaded-PPLN OPA crystal, respectively.}
\end{figure*}

\begin{figure*}
\includegraphics[width=0.7\textwidth]{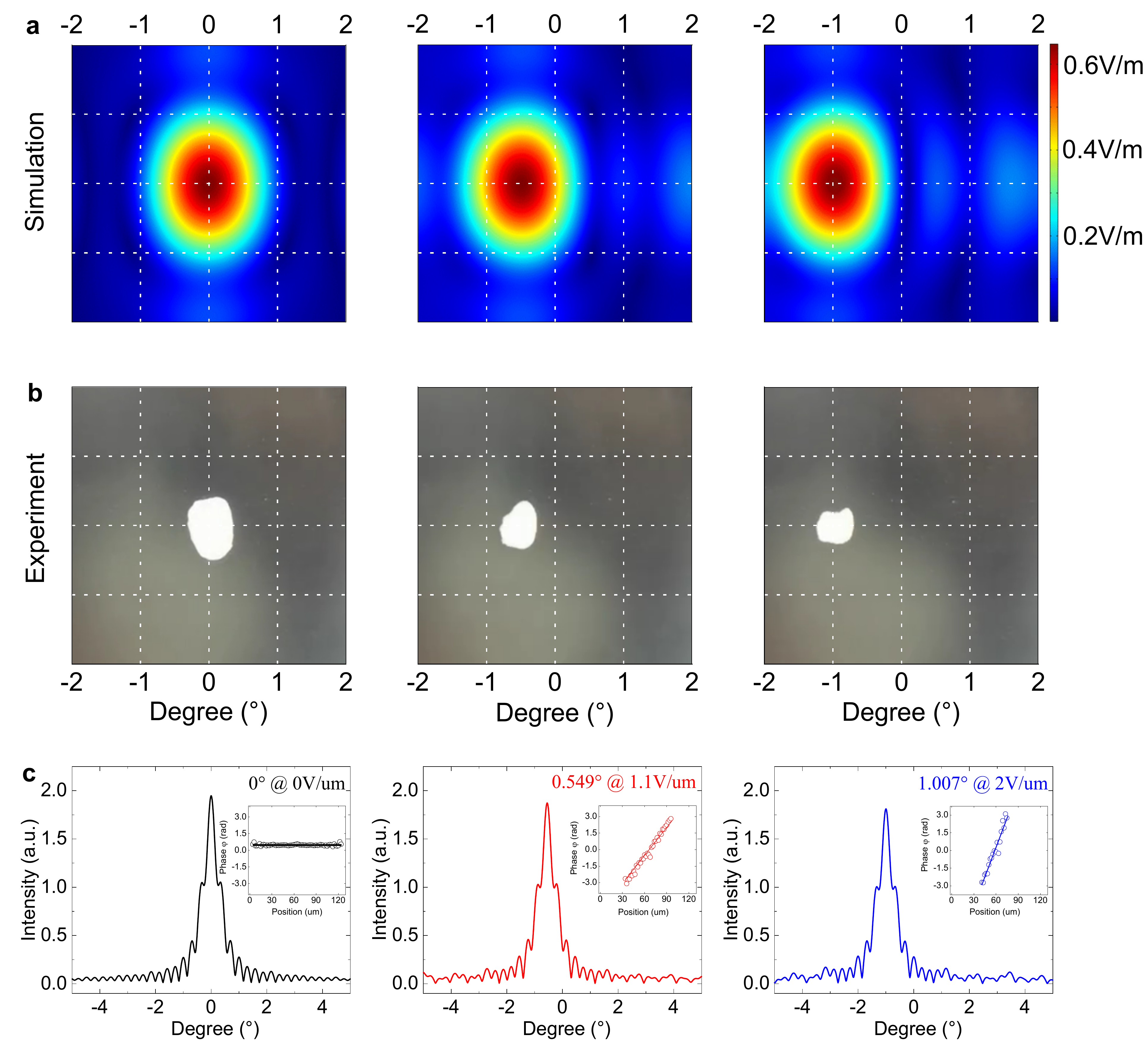}
\caption{\label{6} \textbf{a and b}, The simulated and experimentally obtained far field images of the beam steering shows the one-dimensional angular deflection of the beam. \textbf{c}, The far field patterns of  6-layer cascaded-PPLN OPA crystal  crystal at different voltages of 0V/$\mu$m, 1.1V/$\mu$m and 2V/$\mu$m, respectively, the inset shows the phase distribution curve of the electric field when the beam exits the crystal.}
\end{figure*}

\begin{figure}
\includegraphics[width=0.4\textwidth]{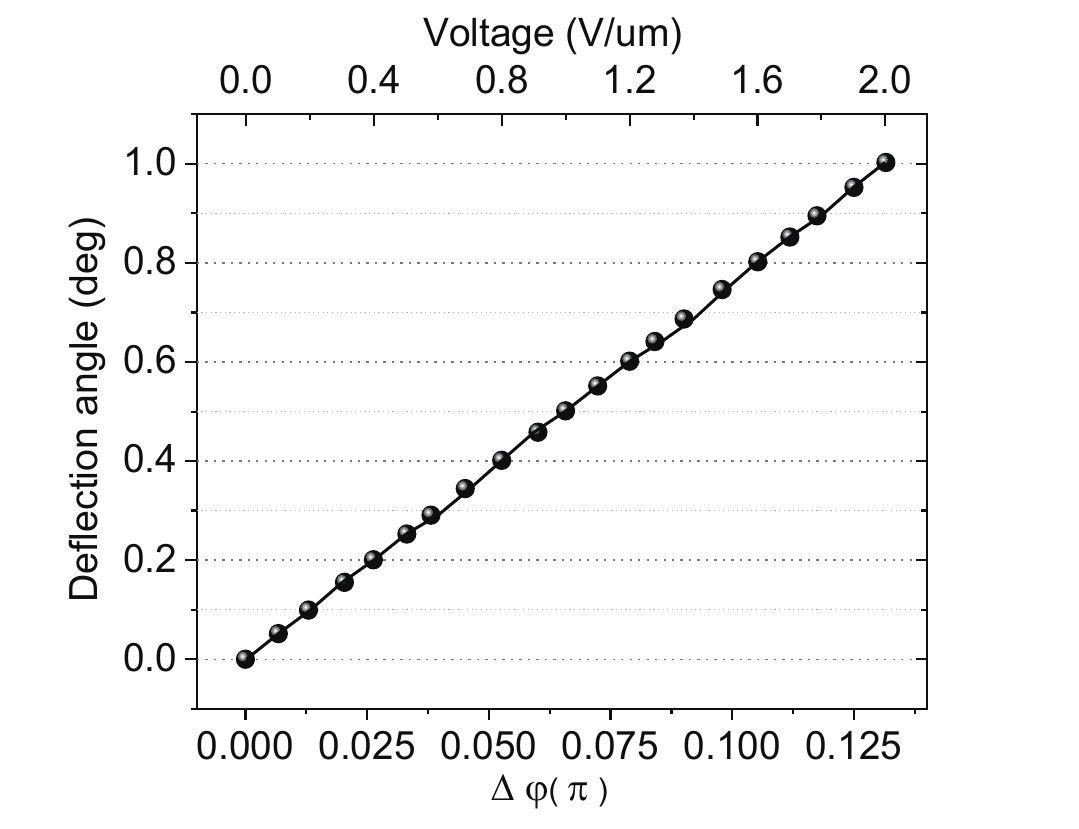}
\caption{\label{7} Beam steering as a function of the electro-optically induced  phase difference $\Delta \varphi $ between adjacent elements. }
\end{figure}

The overall layout of OPA is illustrated schematically in Fig.~\ref{1}a. Its core part is composed of a Z-cut  on-chip cascaded periodically poled lithium niobate (cascaded-PPLN) waveguide, which experiences the largest electro-optic effect, aligns with the external electric field. This waveguide bonded to a SiO$_2$ layer coated LN wafer, where a thick gold layer between the LN wafer and the SiO$_2$ layer served as an electrode, and another electrode is applied on top of the waveguide. A laser input is coupled into the cascaded-PPLN waveguide through an optical fibre, and the output wavefront of the light beam can be controlled by the external applied electric field. The demo structure of a 5-layer cascaded-PPLN waveguide is shown in Fig.~\ref{1}b. The thickness of the waveguide is 50 $\mu$m. For the first cascaded layer (input layer),  the length and width of individual domain set as 3200 $\mu$m and 64 $\mu$m, respectively. As the number of cascaded layers increases, the length and width of individual domains in each layer are, in sequence, reduced by half. Figure~\ref{1}c shows the simulated optical mode profile at $\lambda=$1064 nm, and  Fig.~\ref{1}d shows the simulated electric field distribution in the x-direction at the bottom of the waveguide. One can find that the coupling between the optical modes and the cascaded-PPLN waveguide can be greatly improved in such a way, enabling OPA to achieve a larger deflection angle with a lower voltage drive.

The most important part of controlling beam steering in the principle of one-dimensional OPA is the equal phase difference between each adjacent element when the light exits the array. Its theory is clearly described in the Ref \cite{bib30}. The purpose of the cascaded-PPLN waveguide designed in this paper is to construct an equal phase difference ($\Delta \varphi $) distribution between each adjacent array element, based on electro-optical effect of LN. The deflection angle is determined by ${{\theta }_{s}}=\lambda \Delta \varphi /2\pi d $, where $\lambda $ and $d$ represent the wavelength and the width of single array element respectively, $\Delta \varphi $ is the phase difference between the adjacent array elements. Once $\Delta \varphi $  is changed, the far field pattern will change which means the beam steering could be controlled by $\Delta \varphi $ . Here, we interpret how the equal phase difference  ($\Delta \varphi $) of each adjacent elements is generated in the cascaded-PPLN waveguide. Assuming that the total number of cascaded layers is $N$, $m$ represents the layer of the $m$-th $(0\le m\le N)$, the total number of ferroelectric domains in the $m$-th layer is ${{2}^{m-1}}$. Set ${{x}_{m}}$ to represent the ${{x}_{m}}$-th basic domain in the $m$-th layer. From the first layer (input layer) to the last (output layer),  due to the optical length $l$ of the cascaded layer be halved in turn as the number of cascaded layers increases, the electro-optically introduced phase shift of single ferroelectric domain in each layer is also halved. In addition, it is noticeable that the phase shift in positive ferroelectric domains is $\Delta {{\varphi }_{+}}$ ($\Delta \varphi_\pm =2\pi l\Delta n/\lambda $ , $\Delta n=-n_{e}^{3}{{\gamma }_{33_\pm}}{{E}_{z}}/2$), while in negative domains $\Delta {{\varphi }_{-}}$has an opposite sign because the sign of the electro-optical coefficient $\gamma_{33_\pm}$ is opposite between positive and negative ferroelectric domains. The electro-optically introduced phase shift for single domain in each layer is determined by $\Delta {{\varphi }_{m}}={{2}^{N-m}}\times\Delta {{\varphi }_{N}}$, where $\Delta \varphi {}_{N}$ is the phase shift caused by the last layer and $\Delta {{\varphi }_{m}}$ is the phase shift caused by the $m$-th layer. The total phase shift in the position of the ${{x}_{N}}$-th basic domain could be calculated by
\begin{equation}
\ \Delta {{\varphi }_{{{x}_{N}},tot}}=\sum\limits_{1}^{N}{{{(-1)}^{\left\lceil \frac{{{x}_{m}}}{2} \right\rceil +1}}\times{{2}^{N-m}}\times\Delta {{\varphi }_{N}}} \
\end{equation}
The sign $\left\lceil {} \right\rceil $ represent the ceil function, ${{x}_{m}}=\left\lceil \frac{{{x}_{m+1}}}{{{2}^{N-m}}} \right\rceil $, when $m=N$, ${{x}_{m}}={{x}_{N}}$. $\Delta {{\varphi }_{{{x}_{N}},tot}}$ represents the total phase shift produced after the optical beam passes through the entire cascaded crystal in position of the  ${{x}_{N}}$-th basic domain in the last layer. The phase difference of adjacent basic domain ($\Delta \varphi $) $\Delta {{\varphi }_{{{x}_{N}},tot}}-\Delta {{\varphi }_{x-{{1}_{N}},tot}}$ is always $2\Delta {{\varphi }_{N}}$. Therefore, we can control the continuous variation of $\Delta \varphi $ by applying voltage and thus achieve far-field beam steering.

We performed the simulation of the beam steering capability of the on-chip cascaded-PPLN OPA as shown in Fig~\ref{2}a. The enlarged part is the geometry of cascaded-PPLN waveguide. The blue part and the gray part represent LN with positive and negative ferroelectric domains respectively. It is worth noting that the cascaded-PPLN model used in the simulation has been simplified in order to reduce the computational effort, but the simplification does not affect the actual beam deflection ability of the cascaded-PPLN. The reasons are as follows, the phase shift in single ferroelectric domains is $\Delta {\varphi }_\pm$ ($\Delta \varphi_\pm =2\pi l\Delta n/\lambda $ , $\Delta n=-n_{e}^{3}{{\gamma }_{33_\pm}}{{E}_{z}}/2$) , which means the phase shift caused by single domain in LN is only related to the length of the single domain and the applied voltage. Thus, we reduced the length of the domain and compensate it by increasing the applied voltage in order to reduce the computational effort during the simulation process. Another enlarged part is the electric field distribution when the beam passes through the cascaded-PPLN waveguide without applying voltage. When voltages of 1V/$\mu$m, 2V/$\mu$m, 3V/$\mu$m, 4V/$\mu$m and 5V/$\mu$m were applied to the cascaded-PPLN waveguide, the beam deflection angles were 0.993°, 2°, 3.02°, 4° and 4.991°, respectively (as shown in Fig~\ref{2}a). Figure~\ref{2}b-g shows the far-field patterns at different voltages, where the enlarge part shows the phase curve of the electric field when the beam exits the cascaded-PPLN waveguide. The phase curve is not sloped when no voltage is applied, which means that the far field pattern is not deflected (see  Fig~\ref{2}b). As the applied voltage gradually increases, the phase curve is sloped gradually, which means that the phase difference between adjacent ferroelectric domains also increases and the number of ferroelectric domains required to complete a $2\pi $ phase modulation decreases. As mentioned above, the phase difference between adjacent basic ferroelectric domain $\Delta \varphi $ is always $2\Delta {{\varphi }_{N}}$, in cascaded-PPLN waveguide, $\Delta {{\varphi }_{N}}$ represents the phase shift provided by single domain in the fifth cascade (output) layer. The phase differences between adjacent array elements were calculated to be $0.1315\pi $, $0.2631\pi $, $0.3947\pi $, $0.5262\pi $ and $0.6578\pi $ when voltages of 1V/$\mu$m, 2V/$\mu$m, 3V/$\mu$m, 4V/$\mu$m and 5V/$\mu$m were applied respectively (as shown in Fig~\ref{2}b-g). As know from OPA theory, any emitter pitch spacing much greater than half the wavelength in free-space leads to the presence of grating side lobes. In the cascaded-PLLN waveguide, the width of single domain in the last cascaded layer is set to be 4 $\mu$m, which is larger than half the wavelength. Therefore, the appearance of the grating side lobes is inevitable. The position of the grating side lobes is determined by $\sin \theta =\pm m\lambda /d+\sin {{\theta }_{s}}$ . 
According to the equation ${{\theta }_{s}}=\lambda \Delta \varphi /2\pi d$, the deflection angle can be further improved by reducing $d$ or increasing $\Delta \varphi $ at a certain voltage drive. As shown in Fig.~\ref{3}a, the deflection angle increases form 4.991° to 6.554° and the phase difference between adjacent single domain $\Delta \varphi $ is unchanged, as $d$ is reduced from 4 $\mu$m to 3 $\mu$m. Since the width of single domain is larger than half wavelength, the effect of grating side lobes is still unavoidable. Although the current poling process cannot achieve the width of single domain less than half 1064 nm wavelength, an unequally spaced OPA arrangement can also realize the effect of compressed grating side lobes. In the cascaded-PPLN waveguide, we can arrange the unequal widths of single domains in the last cascaded layer to compress the grating side lobes.

It is well known that the only way to improve the trade off between steering range and angular resolution is increasing the number of control elements, which complicates the control system in the most OPAs but not in cascaded-PPLN OPA. All the array elements are manipulated by only one control unit, the number of array elements is only related to the number of cascade layers, and the number of array elements doubles for each additional cascaded-layer. The divergence angles is inversely proportional to the number of array elements and the full-width at half-maximum (FWMH) divergence profiles of the different cascaded layers PPLN OPA waveguide were simulated in the far-field as shown in Fig.~\ref{4}. It can be observed that the devergence angles exhibit obvious reduction with the the cascaded-layer change from 5 to 8. The beam divergences of 0.159° in a steering range of ±5°was achieved without additional phase tuning units. As the number of cascaded-layer increases, the divergence angle could be further reduced, depending on the advancement of the lithium niobate poled process. Compared with other high-resolution OPA\cite{bib31,bib32,bib33,bib34,bib35}, the present cascaded-PPLN OPA shows great superiority in terms of the number of phase tuning units.

\subsection{Experimental demonstration}

Considering the processing cycle of the cascaded-PPLN OPA waveguide, which is being processed,  here we report a 6-layer cascaded-PPLN in bulk material to verify the feasibility. The fabrication process of the 6-layer bulk cascaded-PPLN is as follows: Firstly, the photoresist is spin-coated onto the MgO:LiNbO$_3$ crystal surface. Secondly, the cascaded pattern is prepared by exposing the crystal surface after drying, and then a layer of metal electrodes is plated on the surface and the cascaded metal electrodes could be formed on the surface after removing photoresist with developer. Finally, a 6-layer cascaded-PPLN crystal was prepared by using voltage pulse application method. The prepared 6-layer cascaded-PPLN crystal is shown in Fig.~\ref{5}a. The cascaded periodically-poled structure is in the middle part of the crystal and the geometric dimensions have been marked in the geometry schematic (as shown in Fig.~\ref{5}b). The thickness of the crystal is 500 $\mu$m. The invert domain structure inside the crystal is shown in  Fig.~\ref{5}c$_1$-c$_4$, where the 6-layer cascaded periodically-poled structure can be clearly seen. It is apparent that the first layer is single positive ferroelectric domain, and each subsequent layer of positive and negative ferroelectric domains are arranged periodically. As the number of layers increases, the width ($W_m$) and the length ($L_m$) of single positive or negative ferroelectric domains are both reduced by half ( $m$ represents the layer of the $m$-th $(0\le m\le 6)$). The experimental setup is shown in supplementary.

Figures~\ref{6}a and \ref{6}b show the simulated and experimental results of the far-field images, respectively. The simulated far-field images are obtained by using the near-to-far field transformation (NFFT). The electric field distribution in the near field is obtained by simulating the propagation of the electrical field in the 6-layer cascaded-PPLN. The equation of the NFFT is
\begin{equation}
\begin{split}
E(u,v,f)=\frac{-1}{i\lambda f}{{e}^{i2\pi f/\lambda }}{{e}^{-i\pi ({{u}^{2}}+{{v}^{2}})/\lambda f}}  \int \int\limits_\infty^\infty E(x,y,0) \\
{{e}^{-i\pi ({{x}^{2}}+{{y}^{2}})/(\lambda f)}}  {{e}^{i2\pi (xu+yv)/(\lambda f)}}dxdy
\end{split}
\end{equation}
where $E(x,y,0)$ represents the electric field distribution at the exit cascaded-PPLN crystal, $E(u,v,f)$ represents the diffraction electric field at position $f$ . The experimentally obtained dynamically tunable far-field pattern distribution is captured by a charge coupled device (CCD). When no voltage is applied to cascaded-PPLN crystal, no slanted phase profile is formed at the exit interface, the diffraction pattern of the far field does not be deflected. When voltages of 1.1V/$\mu$m and 2V/$\mu$m are applied to cascaded-PPLN, beam steering occurs, as shown in Fig.~\ref{6}a-b. The reason for beam steering is the simultaneous action of the electro-optical effect and the cascaded periodic structure, which slopes the phase profiles of the electrical field at the cascaded-PPLN crystal exit interface makes the far field pattern deflected. The steering angle is determined by the phase difference between adjacent ferroelectric domains $\Delta \varphi $, the wavelength $\lambda $ and the width of the array elements $d$ (the width of single domain $W_6$) together. Since the phase difference between ferroelectric domains is always $2\Delta {{\varphi }_{N}}$ ($\Delta {{\varphi }_{N}}$ is the phase shift introduced by the last layer of an individual domain) in cascaded-PPLN, the steering angle could be only determined by the applied voltage when $\lambda $ and $d$ in a certain value. Figure~\ref{6}c shows the 1D interference pattern in the far field. The steering angles are 0.549° and 1.007° at 1.1V/$\mu$m and 2V/$\mu$m, respectively. The beam steering in the far field is related to the phase difference between adjacent ferroelectric domains, so we analyzed the phase distribution of electrical field at the cascaded-PPLN crystal exit interface, as shown in Fig.~\ref{6}c. When no voltage is applied to cascaded-PPLN crystal, the phases of the electric field at the  cascaded-PPLN crystal exit interface are identical and the phase profile is not tilted. After a voltage of 1.1V/$\mu$m is applied to the cascaded-PPLN crystal, the phase shift caused by the individual element in the last layer is $0.0608\pi $ according to $\Delta n=-n_{e}^{3}{{\gamma }_{33_\pm}}{{E}_{z}}/2$ and $\Delta \varphi_\pm =2\pi L_6\Delta n/\lambda $. Thus, the phase difference $\Delta \varphi $ between adjacent elements  is $0.1216\pi$, which means about 17 array elements are needed to achieve a $2\pi $ phase modulation (see Fig.~\ref{6}c). Meanwhile, the steering angle is 0.549° calculated according to ${{\theta }_{s}}=\lambda \Delta \varphi /2\pi d $. Similarly, when the applied voltage is 2V/$\mu$m, the phase shift caused by the individual elements in the last layer comes to $0.1106\pi $, and the phase difference between adjacent elements is $0.2212\pi $, so only 9 elements are required to achieve a $2\pi $ phase modulation (as shown in Fig~\ref{6}c), and the steering angle is 1.007° with an applied voltage of 2V/$\mu$m. The applied voltage has a direct relationship with the phase difference between adjacent elements and the steering angle. The experimentally observed voltage-dependent beam deflection is shown in Fig~\ref{7}. The continuous and quasi-linear steering between 0-1° is observed. In addition, the continually active phase tunability can be realized in the proposed OPA by demonstrating dynamic beam steering, which is shown in supplementary.

One of the most important aspects for current optical phased array technology is its speed. Our cascaded-PPLN OPA has the extremely short response times of the utilized Pockels effect, ultrafast steering operation is expected. Although we don’t characterized the frequency response of the optical beam steering for cascaded-PPLN OPA, the pockels effects have a fast response time of the refractive-index change in the order of 100$fs$ or less\cite{bib36,bib38}. In practice, a modulation speed of 100GHz has experimentally been demonstrated for electro-optical polymer Mach-Zehnder modulators\cite{bib37}, and an electro-optic OPA beam deflectors with a high steering speed of 18GHz has been demonstrated\cite{bib22},
which means the cascaded-PPLN OPA has the ability to achieve high-speed communication.

\section{Conclusion}

In summary, we have proposed a new cascaded-PPLN OPA structure. It could achieve high-speed and continuous beam steering by only one control electronics unit. The beam steering principle behind this structure is the multiple layers cascaded periodic arrangement of ferroelectric domains (positive and negative domains) in the LN crystal. The different Pockels coefficient in the positive and negative ferroelectric domains are used to modulate the phase of the light passing through.  Therefore, the light exhibits an equal phase difference distribution when it is emitted from adjacent domains, and the phase difference of light between adjacent domains can be controlled by applying different voltages to achieve far-field beam steering. We have designed a 5-layer on-chip cascaded-PPLN OPA waveguide with steering angle of 4.991° at voltage of 5V/$\mu$m by simulating. The simulation show that the more wide angle could be achieved when the width of single ferroelectric domain further decreasing with a maturing technology. The beam divergences of 0.159° in a steering range of ±5°was achieved with only one phase tuning unit, which means cascaded-PPLN OPA has potential for high-resolution beam steering and enables new applications in free-space optical communications. Although the device of 5-layer on-chip cascaded-PPLN OPA is not shown in the article, the 6-layer cascaded-PPLN OPA is fabricated in bulk LN and its observed capability of beam steering is agreed well with simulation results. We believe that this new structure provides a new evolution for the realization of highly integrated, high-speed, high-resolution,low-power consumption OPA beam steering systems.

\section{Supplementary material}

\section{acknowledgments}

Shaanxi Key Research and Development Project (Grant No. 2019ZDLGY09-10); Key Innovation Team of Shaanxi Province (Grant No. 2018TD-024); 111 Project of China (Grant No. B14040).

\section{ Author declarations}

\textbf{Conflict of Interest}

The authors have no conflicts to disclose

\section{Data avaliability}

The data that support the findings of this study are available
from the corresponding author upon reasonable request.

\section{References}

\nocite{*}
\bibliography{aipsamp}

\end{document}